\documentclass[twocolumn,twoside,5p]{elsarticle}
\usepackage[utf8]{inputenc}
\usepackage[T1]{fontenc}
\usepackage{txfonts}
\usepackage[dvipsnames]{xcolor}

\usepackage{graphicx}
\usepackage{rotating}
\usepackage{xcolor}
\usepackage{textcomp}
\usepackage{ifthen}

\usepackage{placeins}

\usepackage{isotope}

\usepackage{scalefnt}

\usepackage{amsmath}
\usepackage{bm}
\usepackage{xfrac}
\usepackage{braket}

\usepackage{hyperref}

\definecolor{indianred}{rgb}{0.86, 0.08, 0.24}
\definecolor{royalblue}{rgb}{0.25, 0.41, 0.88}
\definecolor{darkorange}{rgb}{1.0, 0.55, 0}
\definecolor{mediumseagreen}{rgb}{0.24, 0.70, 0.44}
\definecolor{purple}{rgb}{0.5, 0, 0.5}
\definecolor{cyan3}{rgb}{0, 0.80, 0.80}

\newcommand{\la}{\langle}
\newcommand{\ra}{\rangle}

\newcommand{\symboltriangleup}[1][black]{{\color{#1}\scalefont{0.9}{\raisebox{1.2ex}{\begin{turn}{180}$\blacktriangledown$\end{turn}}}}}

\newcommand{\symbolbox}[1][black]{ \raisebox{0.2ex}{\color{#1}\scalefont{1.2}$\blacksquare$}}
\newcommand{\symbolcircle}[1][black]{{\color{#1}\scalefont{1}\ding{108}}}

\newcommand{\symbolboxopen}[1][black]{ \raisebox{.2ex}{\color{#1}\scalefont{1.2}$\square$}}
\newcommand{\symbolcircleopen}[1][black]{\raisebox{-.2ex}{\color{#1}\scalefont{2}$\circ$}}

\newcommand{\bluecircle}{{\scalefont{0.9}\symbolcircle[royalblue]}}

\newcommand{\greentriangleup}{{\scalefont{1.2}\symboltriangleup[plot4]}}
\newcommand{\bluecircleopen}{{\scalefont{0.9}\symbolcircleopen[plot2]}}
\newcommand{\redsquareopen}{{\scalefont{0.9}\symbolboxopen[plot1]}}

\newcommand{\redsquare}{{\scalefont{0.9}\symbolbox[plot1]}}

\newcommand{\elem}[2]{\ensuremath{{}^{#2}\text{#1}}}

\definecolor{plot1}{rgb}{0.86, 0.08, 0.24}
\definecolor{plot2}{rgb}{0.25, 0.41, 0.88}
\definecolor{plot3}{rgb}{1.0, 0.55, 0}
\definecolor{plot4}{RGB}{61,153,86}

\newcommand{\psiref}{\vert \psi_\text{ref} \ra }

\begin{document}

\title{Open-Shell Nuclei from No-Core Shell Model with Perturbative Improvement}

\author[cea]{Alexander Tichai}
\ead{alexander.tichai@cea.fr}
\address[cea]{ESNT, CEA Saclay, IRFU/Service de Physique Nucl\'eaire, F-91191 Gif-sur-Yvette, France}

\author[tud]{Eskendr Gebrerufael}
\ead{eskendr.gebrerufael@physik.tu-darmstadt.de}

\author[tud]{Klaus Vobig}
\ead{klaus.vobig@physik.tu-darmstadt.de}

\author[tud]{Robert Roth}
\ead{robert.roth@physik.tu-darmstadt.de}
\address[tud]{Institut f\"ur Kernphysik, Technische Universit\"at Darmstadt, Schlossgartenstr.\ 2, 64289 Darmstadt, Germany}

\date{\today}

\begin{abstract}
We introduce a hybrid many-body approach that combines the flexibility of the No-Core Shell Model (NCSM) with the efficiency of Multi-Configurational Perturbation Theory (MCPT) to compute ground- and excited-state energies in arbitrary open-shell nuclei in large model spaces. The NCSM in small model spaces is used to define a multi-determinantal reference state that contains the most important multi-particle multi-hole correlations and a subsequent second-order MCPT correction is used to capture additional correlation effects from a large model space. We apply this new \emph{ab initio} approach for the calculation of ground-state and excitation energies of even and odd-mass carbon, oxygen, and fluorine isotopes and compare to large-scale NCSM calculations that are computationally much more expensive.
\end{abstract}

\maketitle

\paragraph*{Introduction}

The solution of the nuclear many-body problem with realistic interactions is at the heart of \emph{ab initio} nuclear structure theory. In recent years tremendous progress has been made in the \emph{ab initio} description of nuclear observables, particularly in the regime of medium-mass nuclei beyond the $p$-shell. Innovative approaches like Coupled Cluster (CC) theory~\cite{DeHj04,BaRo07,HaPa10,KoDe04,PiGo09,BiLa14}, In-Medium Similarity Renormalization Group (IM-SRG)~\cite{HeBo13,TsBo11,Mo15,He14,H15}, or Self-Consistent Green's function (SCGF)~\cite{SoCi13,CiBa13} have been established and provide accurate descriptions of ground-states observables. In a previous work we have shown that many-body perturbation theory (MBPT) with  Hartree-Fock single-particle orbitals yields rapidly convergent perturbation series and that low-order partial sums are in agreement with state-of-the-art CC calculations~\cite{Ti16}, thus, adding to the collection of efficient medium-mass methods.

Despite all the progress, the description of fully open-shell medium-mass systems remains a challenge. The aforementioned methods, in their basic formulation, are limited to ground states of nuclei with closed sub-shells. The ground state of these nuclei is dominated by a single Slater determinant that can serve as a reference state for the construction of the fully correlated eigenstate. Several extensions have been developed to expand the range of the single-determinant methods. Isotopes in the vicinity of shell closures can be tackled by equation-of-motion techniques build on the ground state of a neighbouring closed-shell nucleus \cite{KoDe04}. Further away from shell closures, traditional shell-model approaches, build on a closed-shell core and a small valence-space, combined with non-perturbative valence-space interactions derived from either CC~\cite{Ja16} or IM-SRG~\cite{Bo14,Str16} have been used successfully. 

An important step towards a full no-core description of open-shell nuclei with multi-determinantal reference states is the multi-reference formulation of the IM-SRG \cite{H15}. First applications used particle-number projected Hartree-Fock-Bogoliubov reference states for even-mass isotopes in semi-magic chains~\cite{HeRo09,HeBi13,He14a}. Recently, we merged the multi-reference IM-SRG with the No-Core Shell Model (NCSM)~\cite{NaQu09,RoLa11,BarNa13} to address arbitrary even-mass isotopes and excited states \cite{GeVo17}. These methods are powerful and efficient but far from trivial, both, conceptually and algorithmically.

In this paper we present a much simpler approach, a combination of the NCSM in small model spaces with a low-order MBPT correction to capture correlations from a large space. This hybrid method, for the first time, allows to calculate nuclear ground-state and excitation energies for all open-shell systems in large no-core model spaces. After defining the Hamiltonian, we review multi-configurational perturbation theory and discuss the combination with reference states obtained in the NCSM. We then explore the convergence of the perturbative expansion up to high orders to justify low-order truncations. Using second-order perturbative corrections we perform a detailed study of ground-state and excitation energies for carbon and oxygen isotopes and benchmark with large-scale NCSM calculations. Furthermore, we present the first no-core \emph{ab initio} results for the fluorine isotopic chain out to the extremely neutron-rich \elem{F}{31}.

\paragraph*{Nuclear Hamiltonian}

In all following calculations we start from the chiral nucleon-nucleon (NN) interaction at next-to-next-to-next-to leading order by Entem and Machleidt~\cite{EnMa03}. We include a chiral three-nucleon (3N) interaction at next-to-next-to leading order with a local regulator and a three-body cutoff of $\Lambda_{3N}=400 \,\text{MeV}$ ~\cite{Na07,RoBi12}. The Hamiltonian is softened using a Similarity Renormalization Group (SRG) transformation with a flow parameter $\alpha=0.08\,\text{fm}^4$~\cite{BoFu07,HeRo07,RoRe08,RoLa11,JuMa13}. This transformation induces many-nucleon forces that are included consistently up to the 3N level, many-body forces beyond that level are neglected. This SRG-evolved interaction has been used in a number of calculations in the medium-mass regime \cite{BiLa14,Ti16,GeVo17,HeBi13,He14a} and is, thus, ideally suited to benchmark the present approach.

\paragraph*{Multi-Configurational Perturbation Theory}

The heart of Rayleigh-Schr\"odinger perturbation theory is the definition of an additive splitting, called partitioning, of the nuclear Hamiltonian into an unperturbed part $H_0$ and a perturbation $W$, such that $H = H_0 + W$. 
While the choice of partitioning is simple in the case of standard MBPT with respect to a single Slater determinant, there is no canonical generalization to multi-configurational reference states and several formulations are possible. We adopt the so-called multi-configurational perturbation theory (MCPT) discussed in Refs.~\cite{RoSz03,SuSz04} and also used in Ref.~\cite{Roth09}.

We choose our multi-configurational reference state $\psiref$ to be a normalized eigenvector obtained in a prior NCSM calculation in a model space $\mathcal{M}_{\text{ref}}$,
\begin{align}
\psiref \equiv \sum_{\nu \in \mathcal{M}_\text{ref}} c_\nu \, \vert \phi_\nu \rangle \;, 
\end{align}
where $c_\nu$ denotes the expansion coefficients and $|\phi_\nu\rangle$ the orthonormal many-body basis states, the simple Slater determinant basis of the NCSM in our case. The unperturbed Hamiltonian is chosen such that the reference state fulfills an eigenvalue relation
\begin{align}
H_0\; \psiref = E_\text{ref}^{(0)}\; \psiref \;.
\end{align}
Formally, the unperturbed Hamiltonian can then be written in the spectral representation
\begin{align}
H_0 = E_\text{ref}^{(0)}\; \psiref \la \psi_\text{ref} \vert + \sum_{\nu \notin \mathcal{M}_\text{ref}} E^{(0)}_\nu\; \vert \phi_\nu \ra \la \phi_\nu \vert \;.
\label{Hzero}
\end{align}
Note that only the reference state and not the other eigenstates of the initial NCSM calculation in $\mathcal{M}_{\text{ref}}$ are relevant here.

Following the M{\o}ller-Plesset idea, the zeroth-order energies $E^{(0)}_\nu$ of the unperturbed many-body states $|\phi_\nu \rangle$ outside the reference space, $\nu \notin \mathcal{M}_\text{ref}$, are given by the sum 
$E_{\nu}^{(0)} = \sum_p \epsilon_p$ of single-particle energies $\epsilon_p$ for the states occupied in $|\phi_\nu \rangle$. The single-particle energies are defined via
\begin{align}
\epsilon_{p} \equiv \la p \vert H^{[1]}\vert p \ra + \sum_{rs} \langle pr \vert H^{[2]}\vert p s \rangle\; \gamma_{rs} \;,
\end{align}
where $H^{[1]}, H^{[2]}$ are the one- and two-body parts of the full Hamiltonian, respectively, and $\gamma_{rs}$ is the one-body density matrix of the reference state. In principle, an explicit three-body term can be included as well, however, for the sake of computational simplicity we will later-on use a normal-ordered two-body (NO2B) approximation for the inclusion of 3N interactions~\cite{Geb16}. The zeroth-order reference energy is also defined via these single-particle energies taking into account the multi-determinantal character of the reference state through the mean occupation numbers, i.e., the diagonal elements of the one-body density matrix $\gamma_{pp}$, so that $E_\text{ref}^{(0)} = \sum_p \epsilon_p \gamma_{pp}$.

\begin{figure}[t]
\centering
\includegraphics[width=1\columnwidth]{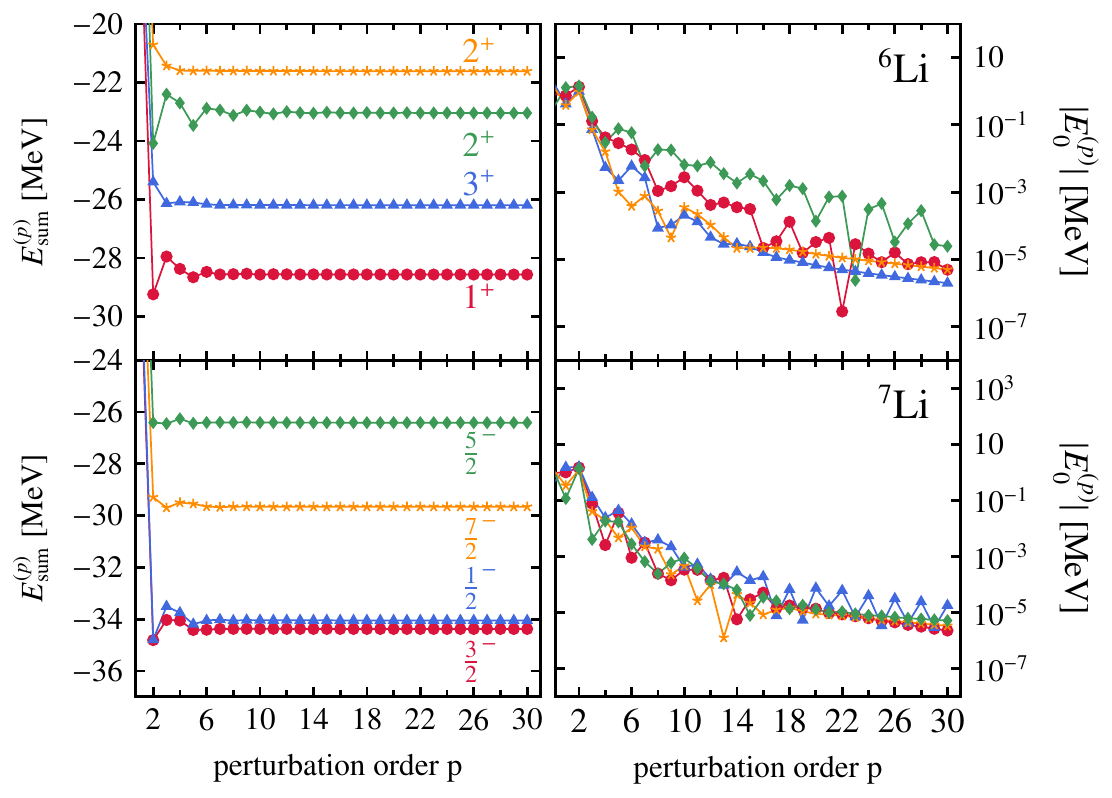} 
\caption{Partial sums (left panel) of $\elem{Li}{6}$ and $\elem{Li}{7}$ for the chiral NN+3N interaction with $\alpha=0.08\,\text{fm}^4$ and truncation parameters $N_{\text{max}}=4$. The corresponding energy corrections for each order are displayed in the right panel, respectively.  All calculations are performed with a harmonic-oscillator basis at $\hbar \Omega =20\,\text{MeV}$.}
\label{highorderplot}
\end{figure}

With the partitioning defined in Eq.~\eqref{Hzero} the zeroth- and first-order contributions to the perturbation series for the energy read
\begin{align}
E^{(0)} &= \la \psi_\text{ref}\vert H_0 \psiref  = E_\text{ref}^{(0)} \;, \\
E^{(1)} &= \la \psi_\text{ref}\vert W \psiref  = \la \psi_\text{ref}\vert H \psiref - E_\text{ref}^{(0)} \;.
\end{align}
Obviously, the sum $E^{(0)}+E^{(1)}$ reproduces the full reference energy, i.e., the eigenvalue obtained for the reference state with the full Hamiltonian $H$ in $\mathcal{M}_{\text{ref}}$.
 
The second-order energy correction has the well-known form
\begin{align}
E^{(2)} 
&= -\sum_{\nu \notin \mathcal{M}_{\text{ref}}} \frac{\vert \langle \psi_{\text{ref}} \vert W \vert \phi_\nu\rangle \vert^2}{E^{(0)}_\nu - E_{\text{ref}}^{(0)} } 
= -\sum_{\nu \notin \mathcal{M}_{\text{ref}}} \frac{\vert \langle \psi_{\text{ref}} \vert H \vert \phi_\nu\rangle \vert^2}{E^{(0)}_\nu - E_{\text{ref}}^{(0)} } \\
&= -\sum_{\mu' \in \mathcal{M}_{\text{ref}}} c_{\mu^\prime} 
\sum_{\mu \in \mathcal{M}_{\text{ref}}} c_\mu^\star \sum_{\nu \notin \mathcal{M}_{\text{ref}}} \frac{\langle \phi_\mu  \vert H \vert \phi_\nu\rangle\langle \phi_\nu \vert H \vert\phi_{\mu^\prime}\ra}{E^{(0)}_\nu - E_{\text{ref}}^{(0)}} \;.
\label{secondorder}
\end{align}
Higher-order perturbative contributions can be formulated in a straight forward manner with the same basic structures for matrix elements and energy denominators. Likewise the perturbative corrections to the many-body states can be evaluated, which is of interest for the computation of observables other than the energies. One can also employ a recursive formulation, as discussed in Refs.~\cite{RoLa10,LaRo12}, to systematically extract high-order corrections.

\begin{figure*}[t!]
\centering
\includegraphics[width=\textwidth]{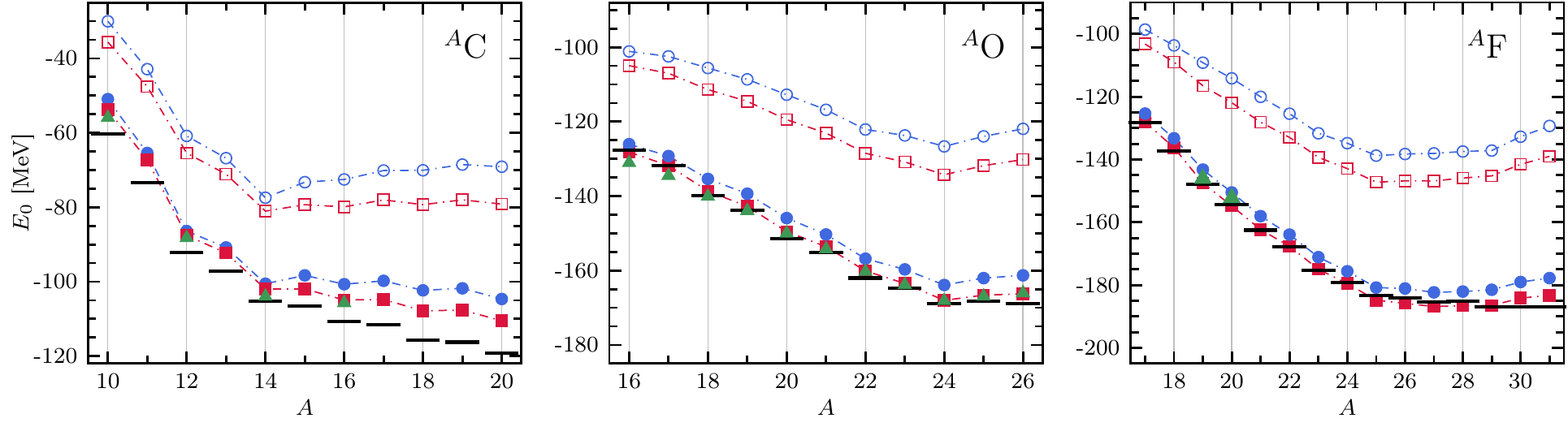}
\caption{Reference energies (\bluecircleopen,\redsquareopen) and second-order NCSM-PT energies (\bluecircle,\redsquare) with $N_\text{max}^{\text{ref}}=0$ and $2$, respectively, for the ground states of $\elem{C}{11-20}$, $\elem{O}{16-26}$, and $\elem{F}{17-31}$ for the NN+3N-full interaction with $\alpha=0.08\,\text{fm}^4$ and model-space truncation $e_{\text{max}}=12$. All calculations are performed with a Hartree-Fock optimized single-particle basis at $\hbar \Omega =20\,\text{MeV}$. Importance-truncated NCSM calculations (\greentriangleup) are shown for comparison~\cite{TiMu18}. Experimental values are indicated by black bars~\cite{AME16}.} 
\label{fig:CFO}
\end{figure*}

The matrix elements in the final expressions for the perturbative corrections only involve the simple unperturbed basis states $\vert \phi_\nu\rangle$, i.e., the Slater determinants that are the basis in $m$-scheme NCSM calculations. Those matrix elements can be readily evaluated using standard NCSM technology.  As an efficient alternative, we employ normal-ordering techniques for evaluating the matrix elements in Eq.~\eqref{secondorder}. We normal order the Hamiltonian with respect to the rightmost determinant $\vert \phi_{\mu^\prime} \ra$ and we redo the normal ordering for each element of the $\mu'$ summation. Similar techniques have been applied in quantum chemistry \cite{HoKa79,HoKa80}. The computational scaling of the second-order correction for large reference spaces is given by $\dim(\mathcal{M}_\text{ref})^2 \cdot n_p^2 \cdot n_h$, where $n_p,n_h$ denote the number of particle and hole states, respectively.

\paragraph*{Combining NCSM and MCPT}

Using the multi-configurational formulation of perturbation theory, we can define a two-stage hybrid approach for the \emph{ab initio} calculation of ground-state energies and excitation spectra. 

The first step consists of an NCSM calculation in a small model space, typically $N_{\max}^{\text{ref}}=0, 2$ or $4$, for a set of low-lying eigenstates. These eigenstates guarantee good $J^{\pi}$ quantum numbers and already contain the most important multi-particle multi-hole correlations as a seed for the perturbative improvement. The second step then consists of the evaluation of the perturbative corrections in a large model space, typically we use a truncation of the single-particle harmonic oscillator basis at $e_{\max}=(2n+l)_{\max}=12$. Each NCSM eigenstate of interest serves as reference state for separate evaluations of the perturbative correction. Thus, perturbation theory is used as a convergence booster that efficiently accounts for correlations from a huge model space.    

Formally, this NCSM-PT approach is guaranteed to converge to the exact result in two different limits: In the limit $N_\text{max}^{\text{ref}} \rightarrow \infty$ the perturbative energy corrections go to zero and we obtain the exact eigenvalue. Alternatively, in the limit of the perturbative order $p\rightarrow \infty$  the exact value is also reproduced for all $N_\text{max}^{\text{ref}}$, provided that the perturbation series converges. In practice, we will restrict ourselves to second-order perturbative correlations, to keep the computational cost at a minimum, and vary $N_{\max}^{\text{ref}}$ to explore the stability of the NCSM-PT energies.

As input for the perturbative corrections the initial three-body Hamiltonian is normal-ordered with respect to the multi-configurational NCSM reference state. Subsequently, we discard the residual three-body part and work with the chiral Hamiltonian in normal-ordered two-body approximation~\cite{Geb16}.

\paragraph*{Convergence Characteristics}

To demonstrate that the multi-configurational perturbation series is well behaved for NCSM reference states, we explicitly evaluate high-order energy correlations adopting the  recursive scheme discussed in \cite{RoLa10,LaRo12}. As benchmark systems we choose $\elem{Li}{6}$ and $\elem{Li}{7}$ using a $N_{\max}^{\text{ref}}=0$ reference space and a small $N_{\max}$-truncated space for the perturbative corrections, so that a direct comparison with explicit NCSM calculations for the same $N_{\max}$ is possible. We use an underlying harmonic-oscillator single-particle basis for these studies. Figure \ref{highorderplot} shows the $p$-th order partial sums in the left-hand panels and the size of the individual perturbative corrections on a logarithmic scale in the right-hand panels. The different data sets correspond to the lowest four eigenstates from the $N_{\max}^{\text{ref}}=0$ space used as reference states in the perturbative calculation. For all states the partial sums converge quickly and higher-order energy corrections are exponentially suppressed. The high-order partial sums agree within a few ten keV with the results of direct NCSM calculations in the same model space. We also find that the low-order partial sums provide a reasonable approximation to the converged value. 
Note that the high-order treatment requires the storage of the many-body basis, and is, therefore, not applicable to medium-mass systems or large model spaces. It serves as a proof-of-principle calculation for the convergence of the perturbation expansion.

\paragraph*{Ground-State Energies}

For heavier systems and larger model spaces, where we cannot compute the perturbation series up to high orders explicitly, we limit ourselves to the computationally simple second-order perturbative correction. We explore ground and excited states through the carbon and oxygen isotopic chains, including even and odd-mass isotopes. For some of these systems, large-scale calculations with the importance-truncated NCSM are still feasible, so that we can benchmark the NCSM-PT results directly. The importance-truncated NCSM calculations within the NO2B approximation are performed up to $N_{\max}=10$ using an optimized natural orbital single-particle basis obtained from diagonalizing a MBPT-corrected one-body density. The use of such natural orbitals improves the model-space convergence and eliminates the dependence on the underlying oscillator frequency~\cite{TiMu18}.

\begin{figure*}[t!]
\centering
\includegraphics[width=\textwidth]{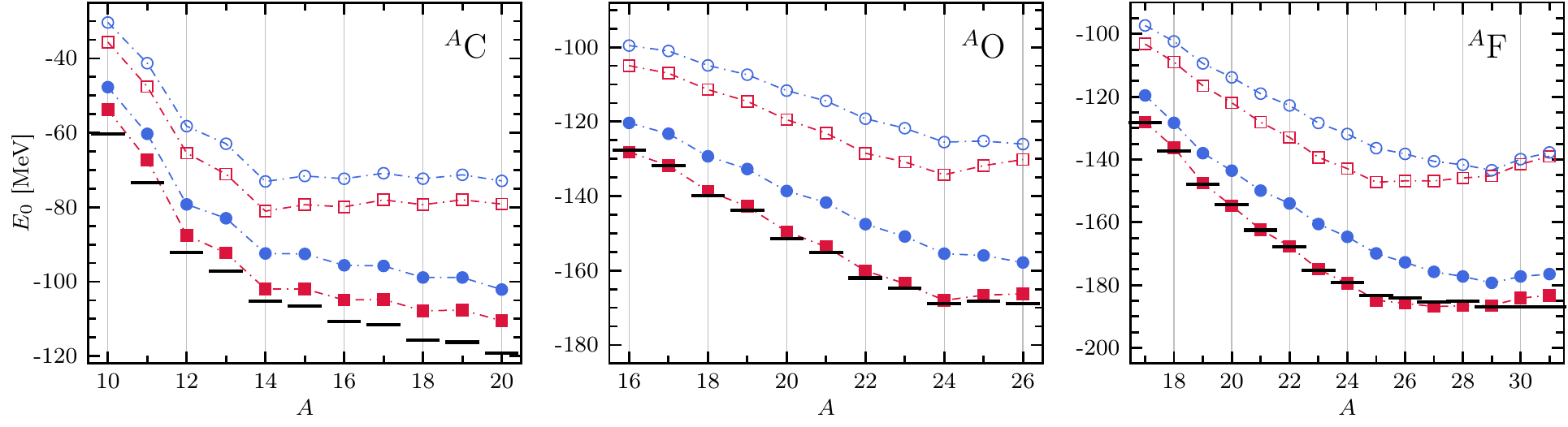}
\caption{Reference energies (\bluecircleopen,\redsquareopen) and second-order NCSM-PT energies (\bluecircle,\redsquare) with $N_\text{max}^{\text{ref}}=2$ for the ground states of $\elem{C}{11-20}$, $\elem{O}{16-26}$, and $\elem{F}{17-31}$ for the NN+3N-ind (squares) and NN+3N-full interaction (circles). The SRG flow parameter is given by $\alpha=0.08\,\text{fm}^4$ and all calculations are performed within a $e_{\text{max}}=12$ truncated model space. We use Hartree-Fock optimized single-particle basis at $\hbar \Omega =20\,\text{MeV}$. Experimental values are indicated by black bars~\cite{AME16}.} \label{fig:CFOintcomp}
\end{figure*}

\begin{figure*}[t!]
\centering
\includegraphics[width=1.0 \textwidth]{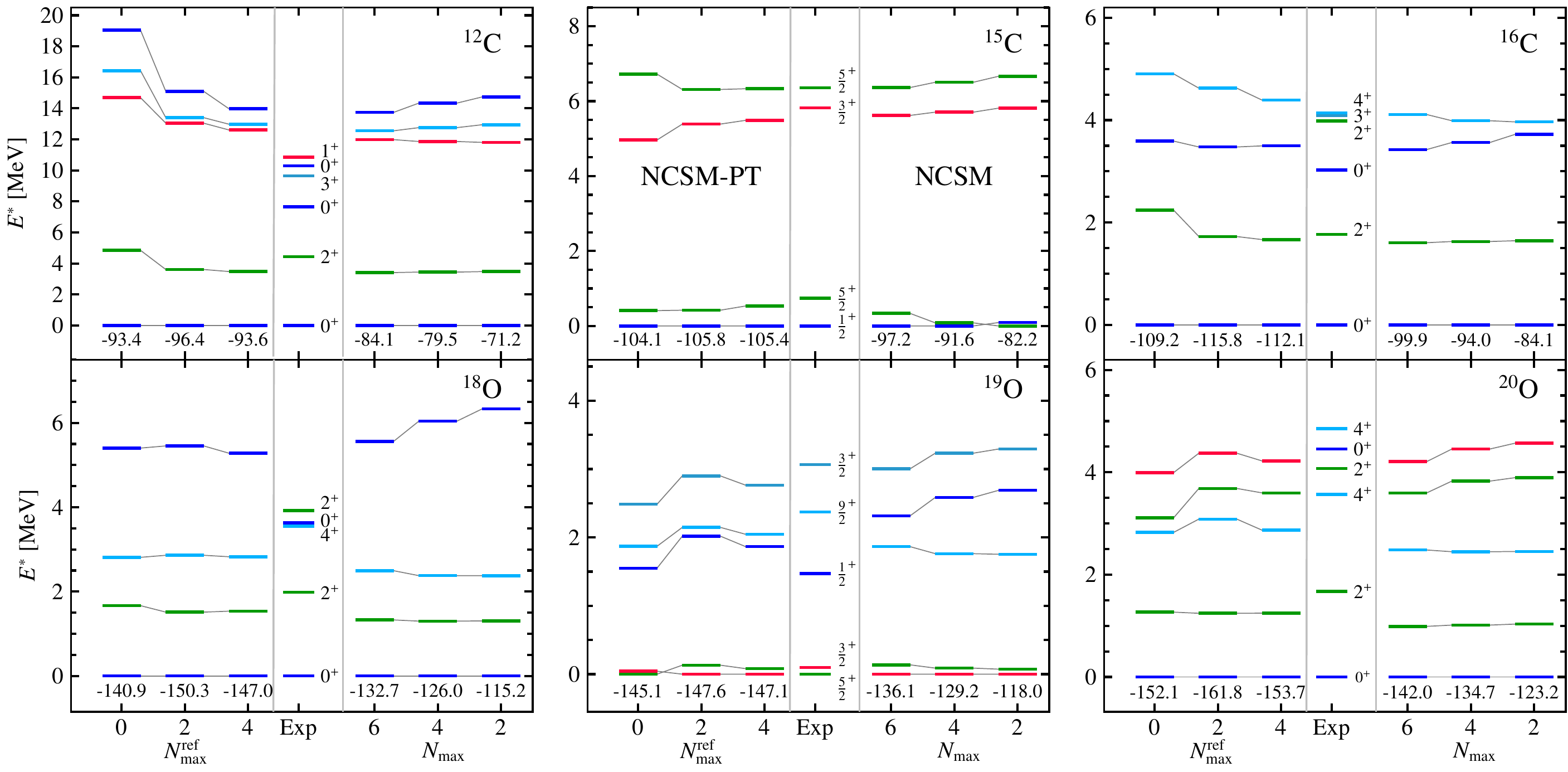}
\caption{Spectra obtained via second-order NCSM-PT for selected carbon and oxygen isotopes for the NN+3N-full interaction with $\alpha=0.08\,\text{fm}^4$ and truncation parameter $e_{\text{max}}=12$. These calculations are performed with a harmonic-oscillator basis at $\hbar \Omega =16\,\text{MeV}$ to separate center-of-mass contaminations. Importance-truncated NCSM calculations for a sequence of model spaces are displayed in the right panel. 
For \elem{O}{19,20} and $N_\text{max}^{\text{ref}}=4$ we introduced an additional truncation $c_\mu c_{\mu^\prime} \geq 10^{-6}$ $ (\mu \neq \mu^\prime)$ for the calculation of the second-order energy corrections in NCSM-PT in order to reduce computing time.}
\label{spectrum}
\end{figure*}

The results for the ground-state energies of carbon, oxygen, and fluorine isotopes are summarized in Fig.~\ref{fig:CFO}, respectively. In addition to the NCSM-PT results including the second-order correction for $N_{\max}^{\text{ref}}=0$ and $2$, we also show the reference energy, i.e., the NCSM eigenvalue obtained in the $N_{\max}^{\text{ref}}$ space. For these calculations we use a Hartree-Fock single-particle basis in order to further optimize the reference states. Both the reference energies and second-order partial sums show a sizable dependence on $N_{\max}^{\text{ref}}$. In general when starting from a $N_\text{max}^\text{ref}=2$ reference state NCSM-PT provides better ground-state systematics than $N_\text{max}^\text{ref}=0$ reference states. In particular NCSM-PT at $N_\text{max}^\text{ref}=2$ almost perfectly reproduces the large-scale IT-NCSM results.
This indicates that the $N_{\max}^{\text{ref}}=2$ space adds important correlations to the reference states than cannot be captured by the second-order perturbative correction. 
We conclude that the NCSM-PT with $N_{\max}^{\text{ref}}=2$ generally provides accurate ground-state energies and an ideal compromise between accuracy and computational efficiency. A single such NCSM-PT calculation requires typically two to three orders of magnitude less computing time than the corresponding importance-truncated NCSM calculation. With the present implementation we will be able to perform NCSM-PT calculations with $N_{\max}^{\text{ref}}=2$ up to the calcium isotopes and slightly beyond.

The NCSM-PT ground-state energies in Fig.~\ref{fig:CFO} for the neutron-rich fluorine isotopes out to heaviest known isotope \elem{F}{31} \cite{SaLu99} represent the first no-core \emph{ab initio} calculations of these nuclei. This regime is relevant for the so-called oxygen anomaly \cite{OtSu10}, i.e., the drastic shift of the neutron dripline from the oxygen to the fluorine isotopic chain. Our calculations show practically constant ground-state energies in the range from \elem{F}{25} to \elem{F}{31}, in agreement with experiment. It will be very interesting to explore this phenomenon with a range of chiral NN+3N interactions, to study its robustness and the theoretical uncertainties resulting from the input interaction. 

Next we will explore the impact of chiral three-body forces on the ground-state systematics. Therefore, we compare our prior results for the NN+3N-full Hamiltonian to ground-state energies obtained from using the chiral NN interaction and keeping SRG-induced many-body terms up to the three-body level (NN+3N-ind). Figure~\ref{fig:CFOintcomp} provides a comparison of the two interactions for the carbon, oxygen and fluorine isotopic chains. All calculations are performed with $N_\text{max}^\text{ref}=2$ reference states. We observe that in all cases the second-order results without chiral 3N forces yield less binding and the agreement with experiment is significantly worse. Of particular importance is the impact on the oxygen dripline. While the inclusion of chiral 3N forces provides the neutron dripline at \elem{O}{24}, the NN+3N-ind interaction is unable to reproduce this property and predicts the neutron rich \elem{O}{25,26} to be bound more tightly than \elem{O}{24}. It was already observed in prior calculations that the inclusion of chiral 3N forces is necessary for the correct reproduction of the experimentally observed dripline~\cite{OtSu10,HeBi13}. 

We observe a similar trend in the fluorine chain, where the inclusion of chiral 3N induces a kink in the ground-state binding energies at \elem{F}{25}, i.e., at neutron number $N=16$ as in the neighbouring \elem{O}{24}.
Beyond \elem{F}{25} ground-state energies remain constant up to \elem{F}{30} which corresponds to opening up the $f_{7/2}$ shell.
The NN+3N-ind interaction predicts a completely different behaviour with smoothly decreasing ground-state energies up to \elem{F}{29}, in contradiction to experiment.

\paragraph*{Excitation Spectra}

By evaluating the second-order correction for different reference states extracted from the NCSM spectrum in the $N_{\max}^{\text{ref}}$ space we can address the excited states directly. We obtain the absolute NCSM-PT energies of the excited states from separate calculations of the second-order correction and subsequently subtract the NCSM-PT ground-state energy to extract excitation energies. Figure \ref{spectrum} presents the excitation spectra of selected carbon and oxygen isotopes compared to direct NCSM calculations. 

It is well known that same-parity excitation energies in NCSM converge much faster with $N_{\max}$ than absolute energies. Therefore, many of the NCSM excitation energies shown in the right-hand columns of each panel in Fig.~\ref{spectrum} are already quite stable. The NCSM-PT, which leads to stable absolute energies for the excited states, can hardly improve the convergence of the excitation energies. We find similar stability with respect to $N_{\max}$ and $N_{\max}^{\text{ref}}$ and good agreement for practically all excitation energies. In cases where the level ordering changes in the NCSM at large $N_{\max}$ the NCSM-PT calculations give the correct level ordering right away, examples are lowest two states in \elem{C}{15}, and the third and fourth state in \elem{O}{19}. As for the ground-state energies, an NCSM-PT calculation for $N_{\max}^{\text{ref}}=2$ provides a good compromise of accuracy and computational efficiency. Due to the scaling with the reference-space dimension, the NCSM-PT calculations with $N_\text{max}^{\text{ref}}=4$ reference states need about two orders of magnitude more computing time than with $N_\text{max}^{\text{ref}}=2$. We further note that absolute energies in NCSM are far from being converged.

\paragraph*{Conclusion and Outlook}

We have introduced a hybrid \emph{ab initio} approach, the NCSM-PT, that combines the flexibility of the NCSM with the efficiency of MBPT techniques to compute ground and excited-state energies in arbitrary open-shell systems in large model spaces. The NCSM in small model spaces is used to define a multi-determinantal reference state that contains the most important multi-particle multi-hole correlations and the second-order correction from multi-configurational perturbation theory are used to capture correlation effects from a large model-space. Everything is formulated in an $m$-scheme basis, so that even and odd-mass nuclei and excited states can be treated directly. We find very good agreement of the ground-state and excitation energies obtained in NCSM-PT with direct NCSM calculations---the accuracy of the NCSM-PT is on par with more demanding approaches like the multi-reference IM-SRG. We presented the first no-core \emph{ab initio} calculations for the of neutron-rich fluorine isotopes, which reproduce the so called oxygen anomaly.

Because of its low computational cost compared to standard NCSM calculations, this approach is ideally suited for exploratory calculations over a large range of nuclei. With the rapid progress in the construction of consistent NN+3N interactions from chiral EFT at various orders and with different regulators \cite{EpKr15,EnMa17}, survey calculations for testing and constraining new nuclear interactions will be of great importance.

\paragraph*{Acknowledgements}

This work is supported by the DFG through contract SFB 1245, the Helmholtz International Center for FAIR within the framework of the LOEWE program launched by the State of Hesse, and the BMBF through contracts 05P15RDFN1 (NuSTAR.DA) and 05P2015 (NuSTAR R\&D). Numerical calculations have been performed at the computing center of the TU Darmstadt (lichtenberg), at the J\"ulich Supercomputing Centre (jureca), at the LOEWE-CSC Frankfurt.


%

\end{document}